%% file: RecSys2024.tex
\documentclass[sigconf, natbib=true]{acmart}
\usepackage{graphicx}
\usepackage{subcaption}
\usepackage{algorithm}
\usepackage{algpseudocode}
\usepackage{enumitem}
\usepackage{multirow}
\usepackage{tabularx}
\usepackage{threeparttable}
\usepackage{color, xspace}

\AtBeginDocument{%
	\providecommand\BibTeX{{%
			\normalfont B\kern-0.5em{\scshape i\kern-0.25em b}\kern-0.8em\TeX}}}
\setcopyright{acmlicensed}
\copyrightyear{2024}
\acmYear{2024}
\setcopyright{acmlicensed}
\acmConference[RecSys '24]{18th ACM Conference
on Recommender Systems}{October 14--18, 2024}{Bari, Italy}
\acmBooktitle{18th ACM Conference on Recommender Systems (RecSys '24),
October 14--18, 2024, Bari, Italy}
\acmDOI{10.1145/3640457.3688117}
\acmISBN{979-8-4007-0505-2/24/10}

\begin{document}
	\title[DNS-Rec: Data-aware Neural Architecture Search for Recommender Systems]{DNS-Rec: Data-aware Neural Architecture Search for Recommender Systems}
	
	\author{Sheng Zhang}
    \authornote{Both authors contributed equally to this work.}
	\affiliation{%
		\institution{City University of Hong Kong}
		\streetaddress{}
		\city{}
		\state{}
		\country{}
		\postcode{}
	}
	\email{szhang844-c@my.cityu.edu.hk}
	
	\author{Maolin Wang}
    \authornotemark[1]
	\affiliation{%
		\institution{City University of Hong Kong}
		\streetaddress{}
		\city{}
		\country{}}
	\email{Morin.wang@my.cityu.edu.hk}

    \author{Yao Zhao}
	\affiliation{%
		\institution{Ant Group}
		\streetaddress{}
		\city{}
		\country{}}
	\email{nanxiao.zy@antgroup.com}

    \author{Chenyi Zhuang}
	\affiliation{%
		\institution{Ant Group}
		\streetaddress{}
		\city{}
		\country{}}
	\email{chenyi.zcy@antgroup.com}

    \author{Jinjie Gu}
	\affiliation{%
		\institution{Ant Group}
		\streetaddress{}
		\city{}
		\country{}}
	\email{jinjie.gujj@antgroup.com}

    \author{Ruocheng Guo}
    \authornote{Ruocheng Guo once worked at CityU and is one of Sheng Zhang and Maolin Wang’s supervisors. This work is not related to ByteDance.}
	\affiliation{%
		\institution{ByteDance Research}
		\streetaddress{}
		\city{}
		\country{}}
	\email{rguo.asu@gmail.com}
	
	\author{Xiangyu Zhao}
    \authornote{Xiangyu Zhao is the corresponding author.}
	\affiliation{%
		\institution{City University of Hong Kong}
		\city{}
		\country{}
	}
	\email{xianzhao@cityu.edu.hk}

    \author{Zijian Zhang}
	\affiliation{%
		\institution{Jilin University \\ City University of Hong Kong}
		\streetaddress{}
		\city{}
		\country{}}
	\email{zhangzj2114@mails.jlu.edu.cn}

    \author{Hongzhi Yin}
	\affiliation{%
		\institution{The University of Queensland}
		\streetaddress{}
		\city{}
		\country{}}
	\email{db.hongzhi@gmail.com}
 
	\renewcommand{\shortauthors}{Sheng Zhang, et.al}

	\begin{abstract}
In the era of data proliferation, efficiently sifting through vast information to extract meaningful insights has become increasingly crucial. 
This paper addresses the computational overhead and resource inefficiency prevalent in existing Sequential Recommender Systems (SRSs). 
We introduce an innovative approach combining pruning methods with advanced model designs. Furthermore, we delve into resource-constrained Neural Architecture Search (NAS), an emerging technique in recommender systems, to optimize models in terms of FLOPs, latency, and energy consumption while maintaining or enhancing accuracy. Our principal contribution is the development of a Data-aware Neural Architecture Search for Recommender System (DNS-Rec). DNS-Rec is specifically designed to tailor compact network architectures for attention-based SRS models, thereby ensuring accuracy retention. It incorporates data-aware gates to enhance the performance of the recommendation network by learning information from historical user-item interactions. Moreover, DNS-Rec employs a dynamic resource constraint strategy, stabilizing the search process and yielding more suitable architectural solutions. We demonstrate the effectiveness of our approach through rigorous experiments conducted on three benchmark datasets, which highlight the superiority of DNS-Rec in SRSs. Our findings set a new standard for future research in efficient and accurate recommendation systems, marking a significant step forward in this rapidly evolving field.

	\end{abstract}

	
	

	\maketitle
    \input{Content/INTRODUCTION}

    \input{Content/METHODOLOGY}
    \input{Content/EXPERIMENT}

    \input{Content/Related_works}
    \input{Content/conclusion}
\begin{acks}
This research was partially supported by Research Impact Fund (No.R1015-23), APRC - CityU New Research Initiatives (No.9610565, Start-up Grant for New Faculty of CityU), CityU - HKIDS Early Career Research Grant (No.9360163), Hong Kong ITC Innovation and Technology Fund Midstream Research Programme for Universities Project (No.ITS/034/22MS), Hong Kong Environmental and Conservation Fund (No. 88/2022), and SIRG - CityU Strategic Interdisciplinary Research Grant (No.7020046), Huawei (Huawei Innovation Research Program), Tencent (CCF-Tencent Open Fund, Tencent Rhino-Bird Focused Research Program), Ant Group (CCF-Ant Research Fund, Ant Group Research Fund), Alibaba (CCF-Alimama Tech Kangaroo Fund (No. 2024002)), CCF-BaiChuan-Ebtech Foundation Model Fund, and Kuaishou.
\end{acks}
    \bibliographystyle{ACM-Reference-Format}
    \bibliography{acmartnew}
    \end{document}

%% file: Content/INTRODUCTION.tex
\section{INTRODUCTION}
	In the face of today's overwhelming data volume, recommender systems have emerged as a powerful tool for capturing users' preferences ~\cite{DL, DL2, mmmlp, Autofield, decision, RL, RL2, RL3, RL4,autodenoise}.
    In particular, sequential recommender systems (SRSs), which model users' historical interactions with items, have become a significant focus in recent research ~\cite{multi-task, Hidasi01, embsearch, TOPN, Context, Contrastive, GNN, AdaFS, automlp} on next-item recommendations. Moreover, attention-based architecture has been proposed to learn the historical interactions of items in attention matrices~\cite{vaswani01}. One representative work of attention-based sequential recommender systems is SASRec~\cite{Kang01}. It employs a multi-head attention mechanism to make predictions based on historical actions from users. Popular variants of SRSs like FDSA~\cite{FDSA}, BERT4Rec~\cite{BERT4Rec} and CORE~\cite{CORE} also display impressive model performance.
    Existing attention-based models, however, incur substantial computational overhead during inference, which leads to inefficient resource utilization and extended inference time.
 
    To mitigate these issues, some network pruning methods ~\cite{Pruning, Pruning2, Structured, Structured2, Structured3} and efficient transformers ~\cite{Linformer, EA, Linrec} have been developed to reduce computational cost. 
    Bridging these developments, recent studies have also incorporated resource-constrained Neural Architecture Search (NAS) to advance the efficiency of recommender systems ~\cite{green,automlp}. This approach entails the incorporation of resource constraints and the automatic selection of smaller, more efficient architectural components, thereby curtailing excessive FLOPs, latency, and energy consumption ~\cite{Proxylessnas, wu2019fbnet}. Models like MONAS ~\cite{hsu2018monas} incorporate operation quantities directly into the reward function using policy-based reinforcement learning. SNAS ~\cite{xie2018snas} models network costs through linear functions, emphasizing the differentiability of resource constraints. RecNAS ~\cite{RecNAS} utilizes annealing and channel pruning to search for a lightweight network architecture.
    
    However, these existing methodologies exhibit certain limitations. Firstly, pruning methods and manually designed efficient models might lead to performance drops if pruning intensities and the network architectures are not properly tuned ~\cite{RecNAS, Linrec, shen2022prune}. 
    Secondly, while current NAS approaches with resource constraints can to a certain degree automatically select appropriate structures,
    they might overly focus on shrinking the model size, thereby resulting in substantial drops in accuracy.

    To address these issues, we propose a novel method, Data-aware Neural Architecture Search for Recommender Systems (DNS-Rec), to tackle the high complexity inherent in attention-based SRS models while simultaneously retaining accuracy, which is based on automatic pruning and the perception of input data batch.
    Firstly, DNS-Rec employs bilevel controllers to conduct automatic pruning, which offers a more dynamic and efficient architectural search paradigm in both depth and width dimensions, compared with manual adjustments. This essential component mitigates the issue of over-parameterized networks, thereby dramatically reducing inference time and storage consumption. Secondly, it introduces data-aware gates to perceive the data environment, and bolsters the overall performance for specific recommendation tasks. It is noteworthy that the employment of data-aware gates enables our network to be adaptively and dynamically generated based on the perception process on the input data batch. 
    Thirdly, it incorporates a linearization operation within the attention mechanism to further curtail computational costs. Finally, DNS-Rec employs a dynamic resource constraint strategy, which not only automatically yields more lightweight structures, but also stabilizes the search process and produces more rational and suitable architectural solutions. 

    We summarize the major contributions of our work as:
    \vspace{-\topsep}
	\begin{itemize}[leftmargin=*]
		\item We propose an automatic pruning method for resource constrained architecture search and apply it to a bilevel controller system, which aims to search for a compact network architecture and enhance the model efficiency.
		\item We design data-aware gates to perceive the data environment, and retain the accuracy of the searched compact network. Information of historical user-item interactions can be perceived through these gates and adapts to deep transformer layers, which enables our network to be dynamically and adaptively generated and improves the performance of the searched network. 
		\item We propose a dynamic resource constraint strategy that adapts to the search stage. It strikes a balance between resource consumption and model accuracy, enabling our architecture search process to be more stable and adaptable.
	\end{itemize}
    \vspace{-\topsep}

%% file: Content/METHODOLOGY.tex
\section{METHODOLOGY}
	\begin{figure*}[t]
        \centering
        \includegraphics[width=0.9\textwidth]{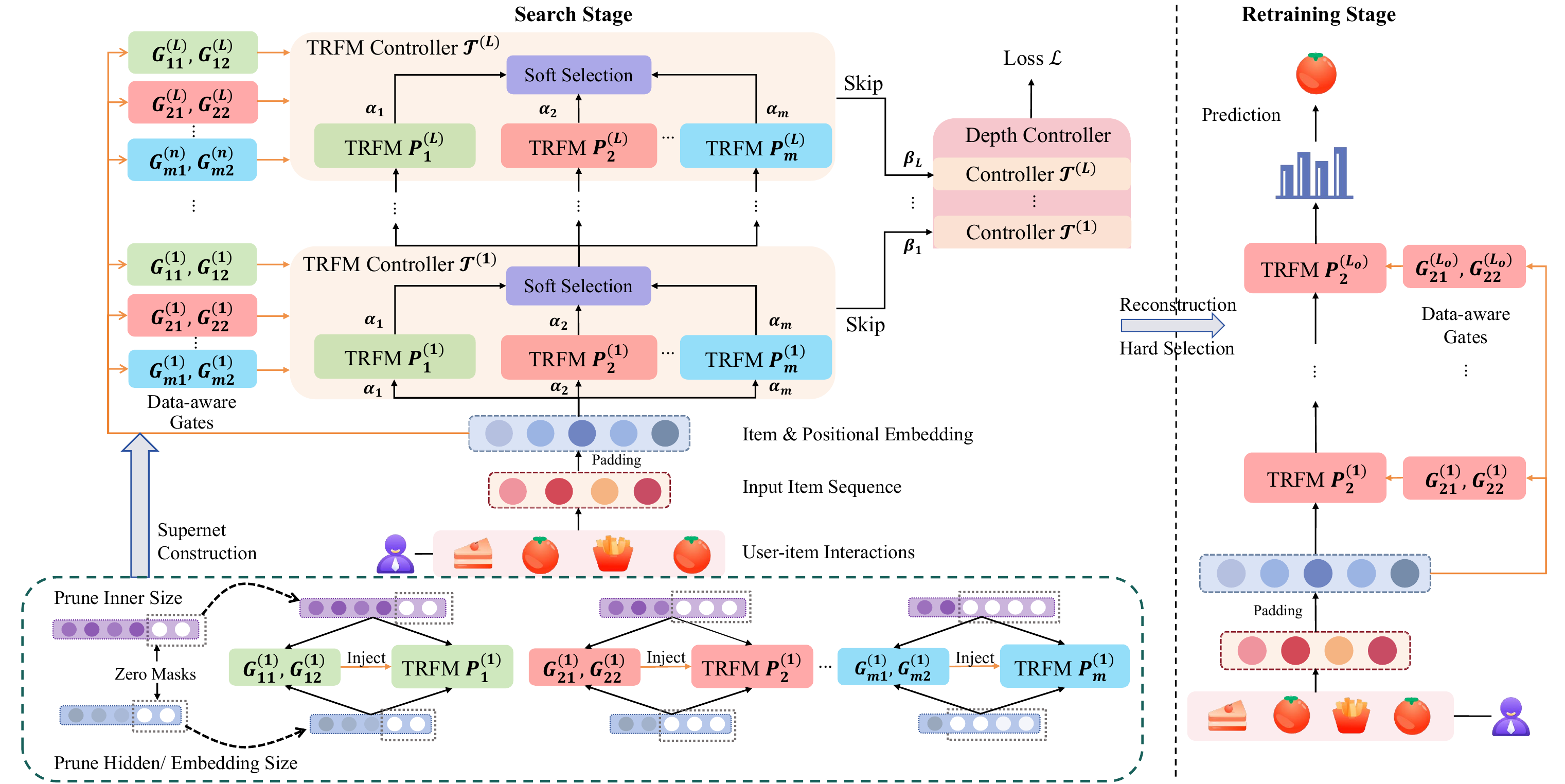}
        \caption{\textbf{Framework of proposed data-aware neural architecture search. 
Our innovative framework 
incorporates a supernet, bilevel controllers, and data-aware gates in a two-stage structure. 
We employ zero masks to prune transformers, and connect each candidate in a certain number of layers for our supernet construction. Architectural weights and model weights are learned in the search process, followed by hard selections for reconstruction and parameter retraining.}}
    \label{fig:network}
    \vspace{-3mm}
\end{figure*}

\subsection{Problem Formulation}
In the attention-based long-term sequential recommendation task, we model a set of users $\mathcal{U} = \{u_1, u_2, \dots, u_{\vert \mathcal{U} \vert}\}$ and items $\mathcal{V}=\{v_1^{(i)}, v_2^{(i)},\dots, v_{\vert \mathcal{V} \vert}^{(i)}\}$. Each user $i$ interacts with a sequence of items $s_i=[v_1^{(i)}, v_2^{(i)}, \dots, v_{n_i}^{(i)}]$. Our network predicts the top-$k$ items with the highest probabilities for recommendations.
To optimize model efficiency with limited resources, we use a differentiable resource constrained NAS (Neural Architecture Search). The search framework, represented as a directed acyclic graph (DAG) $\mathcal{N}=\{x^{(j)}, a^{(i,j)}\vert 1\leq i\leq j\leq n\}$, where $a^{(i,j)}$ is the edge from input node $x^{(i)}$ to output node $x^{(j)}$. Each edge $a^{(i,j)}$ carries out a candidate operation $o^{(i,j)}\in \mathcal{O}=\{ o_1^{(i,j)}, o_2^{(i,j)}, \dots, o_m^{(i,j)}\}$. Following DARTS ~\cite{DARTS}, the architecture weights $\alpha$ can be learned by fusing candidate operations together to generate output node:
\begin{equation}
x^{(j)}=\sum\limits_{k=1}^m\dfrac{\exp(\alpha_k^{(i,j)})}{\sum_{t=1}^m\exp(\alpha_t^{(i,j)})}o_k^{(i,j)}(x^{(i)}).
\end{equation} The objective of constrained NAS is to minimize the loss $\mathcal{L}(x;\alpha,w)$ subject to resource constraints $\mathcal{E}(\alpha)<C$, balancing performance with computational cost. $x$ is the input of the supernet, $\alpha$ are architecture of the supernet, $w$ are model weights, and $\mathcal{L}(\cdot)$ represents the loss function. $\mathcal{E}(\cdot)$ is the function for resource consumption. $C$ is the upper bound of resource constraint.

\subsection{Framework Overview}
Unlike traditional Sequential Recommendation Systems (SRSs) and Neural Architecture Search (NAS) methods that frequently struggle to maintain a balance between accuracy and resource utilization, our novel NAS framework, tailored for resource-constrained environments in long-term SRSs, substantially enhances accuracy while decreasing inference time. 
This framework is comprised of three principal components and structured into two distinct stages.



As illustrated in Figure \ref{fig:network}, our framework utilizes a \textbf{supernet} during the architecture search phase, and applies zero masks to selectively prune transformers. Each transformer is composed of multiple layers, serving as a fundamental unit for evaluating diverse network configurations. The \textbf{bilevel controllers} play a pivotal role in optimizing the network architecture by efficiently adjusting both the network's width and depth. Furthermore, the \textbf{data-aware gates} maintain the integrity of the data and modify the networks dynamically. These gates embed user-item interactions into feed-forward networks by leveraging batch-specific data-aware structures.

In the \textbf{search stage}, the data-aware gates and the bilevel controllers collaborate to reduce computational complexity and pinpoint the optimal network structure. Following this, in the \textbf{retraining stage}, the network is reconstructed based on selected transformers and retrained to ensure its effectiveness and adaptability for the specific recommendation task.

\subsection{Supernet Construction}
Existing resource constrained NAS methods construct supernet simply in repeated layers, ignoring the dynamic data environment and the necessity of retaining accuracy ~\cite{RecNAS}.
Uniquely, we design a supernet shown in Figure \ref{fig:network} for data-aware , which comprises three NAS essential components: transformer controller, depth controller and data-aware gates. 
In this section, we will elucidate the functions of these components in our elaborately designed supernet.
\subsubsection{Embedding Layer}
For long-term sequential recommendation, information of both items and their positions should be encoded into the model ~\cite{Embedding}. We denote the length of input user-item interactions as $N$, and embedding size as $d$. For a user $u_i$ who has a interaction sequence $s_i=[v_1, v_2, \dots, v_t, \dots v_{n_i}]$, the $t$-th item $v_t\in \mathbb{R}^{D_t}$ and its position $p_t\in \mathbb{R}^{D_t}$ can be projected into a dense representation $\boldsymbol{e}_t^s$ and $\boldsymbol{e}_t^p$ respectively through embedding layer:
\begin{equation}
    \boldsymbol{e}_t^s = \boldsymbol{W}_t^s v_t, \ \ \boldsymbol{e}_t^p = \boldsymbol{W}_t^p p_t,
\end{equation}
where $\boldsymbol{W}_t^s \in \mathbb{R}^{d\times D_t},  \boldsymbol{W}_t^p \in \mathbb{R}^{d\times D_t}$ are trainable weighted matrices for $t$-th item and positional embedding, and $D_t$ is its corresponding dimension. Finally the user-item interaction can be represented as:
\begin{equation}
    \boldsymbol{E} = [\boldsymbol{e}_1^s+\boldsymbol{e}_1^p, \boldsymbol{e}_2^s+\boldsymbol{e}_2^p, \cdots, \boldsymbol{e}_N^s+\boldsymbol{e}_N^p]^{\mathrm{T}}.
\end{equation}

\subsubsection{Linearization Operation Transformer Layer Construction}
The fundamental idea of the attention layer in transformer structure is learning item interactions using the dot-product of query matrix $\boldsymbol{Q}$ and key matrix $\boldsymbol{K}$, which generates attention scores to learn sequential representations. 
However, the computational complexity of the dot-product of $\boldsymbol{Q}$ and $\boldsymbol{K}$ is high when $N>d$. To reduce the inference time, we adopt a linearization operation that changes the order of the dot-product. We separate $\boldsymbol{Q}$ and $\boldsymbol{K}$ from the softmax operator by implementing $L_2$ norm and elu mapping, which has been proved to be efficient in previous work ~\cite{Linrec}. The adopted linear attention mechanism can be formulated as follows:
\begin{equation}
\boldsymbol{A}'(\boldsymbol{Q,K,V}) = \mathcal{A}_1\left(elu(\boldsymbol{Q})\right) \left(\mathcal{A}_2\left(elu(\boldsymbol{K})\right)^{\mathrm{T}} \boldsymbol{V}\right),
\end{equation}
where $\mathcal{A}_1(\boldsymbol{Q}_r) = \frac{1}{\sqrt{d} \Vert \boldsymbol{Q}_r \Vert_2}\boldsymbol{Q}_r$ and $\mathcal{A}_2(\boldsymbol{K}_c) = \frac{1}{\sqrt{d} \Vert \boldsymbol{K}_c \Vert_2}\boldsymbol{K}_c$ are $L_2$ normalization mappings. $\boldsymbol{Q}_r$ denotes the $r$-th row of $\boldsymbol{Q}$ for $\forall r \in [1,N]$, and $\boldsymbol{K}_c$ represents the $c$-th column of $\boldsymbol{K}$ for $\forall c\in [1,d]$. 

\begin{figure}[t]
    \centering
    \includegraphics[width=0.8\linewidth]{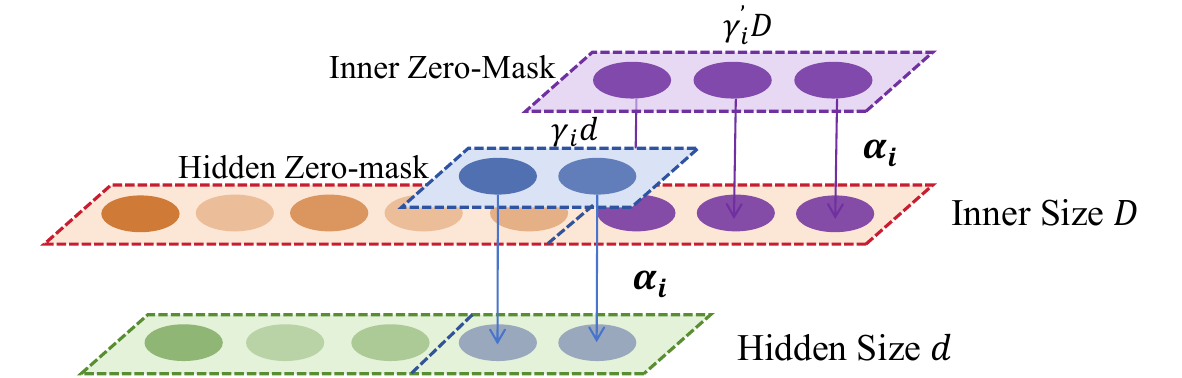}
    \caption{\textbf{Zero-masks designed for pruning hidden and inner size in candidate transformers. 
    Dimensions pruned in hidden and inner layers are replaced by 0 using these masks. }}
    \label{fig:mask}
    \vspace{-5mm}
\end{figure}

To optimize the network architecture, we utilize zero-masks for pruning by implementing $m$ candidate transformers. Each transformer has dimensions that are selectively obscured using zero-masks. We have developed two specific types of zero-masks: a hidden zero-mask targeting the hidden dimensions, and an inner zero-mask for the inner dimensions, both employed to effectively prune the transformers. Each transformer, denoted as $\boldsymbol{P}_i$ for $i \in [1, m]$, is equipped with masks of sizes $\gamma_i d$ and $\gamma_i^{\prime} D$ for its hidden and inner dimensions, respectively. Here, $\gamma_i$ and $\gamma_i^{\prime}$ indicate the proportion of the dimensions that are masked, while $d$ and $D$ signify the consistent dimensions across the supernet. The operation of the zero-mask mechanism is illustrated in Figure~\ref{fig:mask}.
Additionally, to control the depth of the network architecture, each candidate transformer is structured with $L$ layers and incorporates specific data-aware gates, the particulars of which are detailed in Subsection~\ref{data-aware gates}.



\subsection{Bilevel Controllers}
Most existing resource constrained NAS methods search for compact network structures without effective automatic pruning ~\cite{Proxylessnas, RecNAS}.
To tackle this issue, we employ bilevel controllers to search in both width and depth dimensions of the network and apply them to our proposed DNS-Rec framework, which offers a more dynamic and efficient architectural search paradigm.
\subsubsection{Transformer Controller}
The transformer controller, as the primary component of the bilevel controllers, is designed to search for an optimal dimension of the transformer structure with varying pruning intensities. Specifically, the transformer controllers control the architecture weights of candidate transformers with a pruned attention layer, feed-forward network, and data-aware gates. Each candidate transformer together with its data-aware gates carries a weight $\alpha_i$ for $\forall i\in[1, m]$ that represents their shared significance, which can be trained during the search stage. Stacked transformers in $L$ layers share the same $\alpha_i$ in each update so that the dimensions of transformer layers can match with each other.
\subsubsection{Depth Controller}
Although employing redundant transformer layers might rarely have negative impacts on accuracy, it could potentially increase the inference time. Therefore, we propose a depth controller to select the optimal number of transformer layers for our architecture. This results in a final compact network in both width and depth dimensions. As is shown in Figure \ref{fig:network}, after each transformer controller $\boldsymbol{\mathcal{T}}^{(l)}$ generates the output, it has two options, i.e., conveying the output to the next layer of the network, or skipping all the remaining structures to the final prediction layer with trainable weights $\beta_l$ for $\forall l\in[1, L]$.

\subsection{Data-aware Gates}
\label{data-aware gates}
Since the crucial sequential information tends to get weaker as being conveyed to deeper transformer layers, and applying a smaller network might lead to a greater loss of initial information, it is necessary to find a better network design where prior information can be properly learned. For personalized recommendation tasks,~\cite{PEPNet} teaches personalized and author-side features that are separated from general input to the model to enhance final performance. 

\begin{figure}[t]
    \centering
    \includegraphics[width=0.9\linewidth]{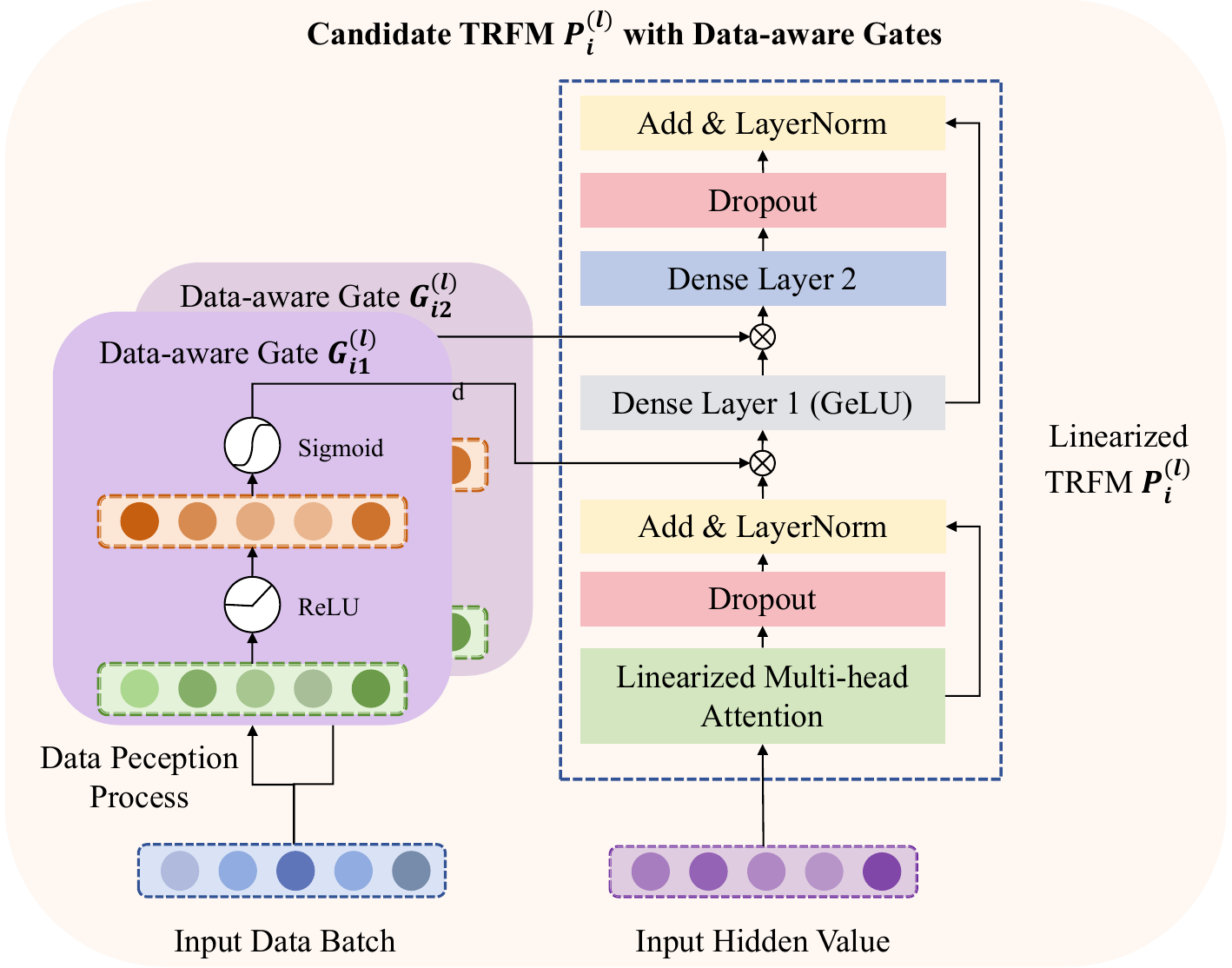}
    \caption{\textbf{Mechanism of data-aware gates. Dual-layer data-aware gates learn input data batch by feature crossing and data perception. Learned information will be adaptively injected into corresponding deep transformer layers.}}
    \label{fig:gate}
    \vspace{-5mm}
\end{figure}

Motivated by this, we design data-aware gates to perceive the input data batch and transmit it to essential positions of transformer layers. The employment of data-aware gates emphasizes the information from input data and enables our network to be adaptively and dynamically generated based on the perception process. The mechanism of our data-aware gates is shown in Figure \ref{fig:gate}, where input data batch $\boldsymbol{X}$ perceived by gates $\boldsymbol{G}_{i1}^{(l)}(\boldsymbol{X})$ and $\boldsymbol{G}_{i2}^{(l)}(\boldsymbol{X})$ are injected into $i$-th candidate linearized transformer $\boldsymbol{P}_i^{(l)}$. Each gate adaptively and independently learns input information to retain the accuracy of the network. In addition, each data-aware gate is bound with a specific dense layer in the feed-forward network, as the input data batch should be differently learned for various sizes of transformers in different depths.

In the data-aware gates $\boldsymbol{G}_{i1}^{(l)}(\boldsymbol{X})$ and $\boldsymbol{G}_{i2}^{(l)}(\boldsymbol{X})$, the input data batch $\boldsymbol{X}$ is transferred for feature crossing in the first layer:
\begin{equation}
    \boldsymbol{X}_k' = \mathrm{ReLU} (\boldsymbol{XW}_{k}^{(1)}+\boldsymbol{b}_{k}^{(1)})), \ \ k=1,2,
\end{equation}
where $\boldsymbol{W}_G^{(i)}$ and $\boldsymbol{b}_G^{(1)}$ are trainable weighted matrix and bias. Then the scores $\boldsymbol{\delta}$ of data perception can be generated in the gates:
\begin{equation}
    \boldsymbol{\delta}_k = \gamma \ast \mathrm{Sigmoid}(\boldsymbol{X}_k'\boldsymbol{W}_{k}^{(2)}+\boldsymbol{b}_{k}^{(2)}),\ \ k=1,2,
\end{equation}
where $\gamma$ is the scaling hyperparameter, which constrains the range of output. We set $\gamma = 2$ in our research. $\boldsymbol{W}_{k}^{(2)}$ and $\boldsymbol{b}_k^{(2)}$ are trainable weighted matrix and bias in the second layer.

We inject learned sequential information into fully connected layers in the feed-forward network of the transformer to enhance input signals adaptively in deep layers. Then we calculate the output $\boldsymbol{T}_i^{(l)}$ of $i$-th candidate transformer $\boldsymbol{P}_i^{(l)}$ in layer $l$ by:
\begin{equation}
    \boldsymbol{T}_i^{(l)}=\boldsymbol{P}_i^{(l)}(\boldsymbol{G}_{i1}^{(l)}(\boldsymbol{X}), \boldsymbol{G}_{i2}^{(l)}(\boldsymbol{X}), \boldsymbol{\mathcal{T}}^{(l-1)}),\ \ l\geq 2
    \label{tr}
\end{equation}
where $\boldsymbol{\mathcal{T}}^{(l-1)}$ is the hidden value of $l-1$-th transformer. The mechanism of feed-forward networks can be organized as follows:
\begin{equation}
    \begin{aligned}
    \boldsymbol{F}_1^{(l)} &= \mathrm{GeLU} ((\boldsymbol{\delta}_1^{(l)} \otimes \boldsymbol{S}^{(l)}) \boldsymbol{W}_{\boldsymbol{F}_1}^{(l)} + \boldsymbol{b}^{(l)}), \\
    \boldsymbol{F}_2^{(l)} &= (\boldsymbol{\delta}_2^{(l)} \otimes \boldsymbol{F}_1^{(l)}) \boldsymbol{W}_{\boldsymbol{F}_2}^{(l)} + \boldsymbol{b}^{(l)}, \\
    \boldsymbol{T}_i^{(1)} &= \mathrm{LayerNorm}(\boldsymbol{S}^{(1)} + \mathrm{Dropout}(\boldsymbol{F}_2^{(l)})), \\
    \boldsymbol{\mathcal{T}}^{(0)} &= \boldsymbol{E}, 
    l\in[1,L], \ \ i\in [1,m],
    \end{aligned}
\end{equation}
where 
$\boldsymbol{S}^{(l)}$ is the multi-head attention matrix and output of $l$-th attention layer $\boldsymbol{F}_1^{(l)}$, $\boldsymbol{F}_2^{(l)}$ are the outputs of two dense layers injected with perceived information in the feed-forward network.

\subsection{Data-aware Architecture Search Stage}
Traditional resource constrained NAS methods lack the ability to generate the most suitable network for a specific recommendation task ~\cite{RecNAS, Proxylessnas}. To tackle this problem, we introduce the data-aware neural architecture search to perceive data environment within bilevel controllers, under dynamic resource constraint and uniquely designed optimization framework.
\subsubsection{Soft Selection}
In each transformer controller, candidates are fused with trainable weights $\alpha_1,\alpha_2, \dots, \alpha_m$. We adopt the Gumbel-softmax trick ~\cite{Gumbel} to conduct soft selection with weight $p_i$, enabling the network to be searched in a differential manner, which follows 
the framework of DARTS~\cite{DARTS} and can be formulated as: 
\begin{equation}
    \begin{gathered}
        p_i = \dfrac{\exp((\log(\alpha_i)+g_i)/\tau)}{\sum\limits_{k=1}^m\exp((\log(\alpha_k)+g_k)/\tau)},\\
        where  \ \ g_k = -\log(-u_k), u_k\sim Uniform(0,1),
    \end{gathered}
\end{equation}
where $p_i$ is the decision weight, and $\tau$ is the temperature hyperparameter that smooths the soft selection. Similarly, the probability of selecting $j$-th candidate skipping operation from $L$ transformer controllers can also be calculated by the trainable weights $\beta_j (\forall j\in [i,L])$ using the Gumbel-softmax trick. Then the output $\boldsymbol{Y}$ of our bilevel controllers becomes:
\begin{equation}
\label{weight}
    \begin{aligned}
        \boldsymbol{Y}& = \sum\limits_{j = 1}^L q_j
        \boldsymbol{\boldsymbol{\mathcal{T}}}^{(l)}= \sum\limits_{j = 1}^L \sum\limits_{i=1}^{m}p_i q_j \boldsymbol{T}_i^{(l)} \\
        & =\sum\limits_{j = 1}^L\sum\limits_{i=1}^{m}p_i q_j \boldsymbol{P}_i^{(l)}(\boldsymbol{G}_{i1}^{(l)}(\boldsymbol{X}), \boldsymbol{G}_{i2}^{(l)}(\boldsymbol{X}), \boldsymbol{\boldsymbol{\mathcal{T}}}^{(l-1)}).
    \end{aligned}
\end{equation}

Similar to the definition of $p$, $q$ in Eq.(\ref{weight}) is the decision weight of the skipping operations, determining the number of transformer layers in the searched network structure. By employing this strategy, we are able to search for a compact architecture with a uniform size and avoid redundant forward propagation. 
Denote that the output of the supernet as $y$, we derive the cross entropy loss function defined below for the search stage:
\begin{equation}
    \mathcal{L}_{CE} = \sum\limits_{i=1}^{n_i} \left( y\log(\hat{y}_i) + (1-y) \log (1-\hat{y}_i) \right)
\end{equation}

\subsubsection{Dynamic Resource Constraint}
Considering that most devices in real-world applications are resource-constrained and that neural networks with over-parameterized architectures may consume excessive energy, we propose a dynamic resource constraint as the penalty term. This aims to facilitate the search for an efficient and effective network architecture. In our research, we use the quantity of floating point operations (FLOPs) to measure the resource consumption. The FLOPs of the searched network structure, denoted as $FLOPs(\cdot)$, is determined by soft selections within bilevel controllers. Specifically, for searchable transformer blocks, we define dynamic resource constraint $\mathcal{L}_E$ as follows:
\begin{equation}
    \mathcal{L}_{RC} = \dfrac{L_{t}}{L}\sum\limits_{j=1}^{L}\sum\limits_{i=1}^m p_i q_j FLOPs(\boldsymbol{T}_i^{(j)}),
    \label{flops}
\end{equation}
where $L_{t}$ represents the number of transformer blocks decided by temporary hard selection in depth controller in $t$-th epoch. 
Due to the considerable differences between candidate pruned transformer layers, the search process can easily become ensnared in the trap of allocating the smallest or largest architecture components with the highest weight, which is often impractical and undermines the stability of architecture search. Therefore, we utilize this dynamic resource constraint strategy adapted to the search conditions in different epochs. This is achieved by conducting hard selections on a number of transformer layers to approximately estimate the FLOPs based on the current state of the supernet. 

Finally, we combine the
cross entropy loss of predictions and resource constraint as the loss functionn $\mathcal{L}$ in the search stage:
\begin{equation}
    \mathcal{L} = \mathcal{L}_{CE}+\lambda\mathcal{L}_{RC},
    \label{loss}
\end{equation}
where $\lambda$ is the tradeoff parameter.

\subsubsection{Optimization}
Based on the aforementioned techniques, the result of the architecture search is mainly determined by 
parameters of transformer controllers $\boldsymbol{\alpha}$,  parameters of depth controller $\boldsymbol{\beta}$. Since jointly learning the three groups of parameters in the same batch of training data by traditional unsupervised learning~\cite{Efficient} may lead to over-fitting~\cite{Autofield}, we train the architecture parameters and model weights separately to address this problem.

Therefore, we use a batch of training data to update model weights, and a batch of data to update architectural parameters, for which we formulate a bi-level optimization problem~\cite{Bilevel}:
\begin{equation}
    \begin{aligned}
        &\min\limits_{\boldsymbol{\alpha}, \boldsymbol{\beta}} \mathcal{L}_{val} (\boldsymbol{\mathcal{W}}^{\ast}(\boldsymbol{\alpha},\boldsymbol{\beta}), \boldsymbol{\alpha},\boldsymbol{\beta})\\
        &\mathrm{s.t.}\ \ \boldsymbol{\mathcal{W}}^{\ast}(\boldsymbol{\alpha},\boldsymbol{\beta}) = \mathop{\arg\min}\limits_{\boldsymbol{\mathcal{W}}} \mathcal{L}_{train} (\boldsymbol{\mathcal{W}}, \boldsymbol{\alpha}^{\ast}, \boldsymbol{\beta}^{\ast}),
    \end{aligned}
    \label{optimization}
\end{equation}
where $\mathcal{L}_{val}$ and $\mathcal{L}_{train}$ are loss functions consistent with $\mathcal{L}$ in Eq.(\ref{loss}).
We initialize the iteration number $t$ as 0, and for each iteration, we first sample a batch of the training dataset, during which the model weights $\boldsymbol{\mathcal{W}}$ can be updated. For every $f$ iterations in one training epoch, we dynamically adjust $L_t$ to impose a resource constraint penalty on the loss function. Then we sample a mini-batch from the validation dataset and update the architectural weights $\lbrace \boldsymbol{\alpha}, \boldsymbol{\beta}\rbrace$ for searching. At the end of each iteration, the number of iteration $t$ is updated.
\subsection{Retraining Stage}
Suboptimal weighted structures of the network derived in the search stage could do harm to the performance of the recommendation. To tackle this problem, in the retraining stage, we reconstruct the architecture with selected transformers and skipping operations, followed by retraining the reconstructed model. 

To alleviate negative impacts from suboptimal structures, 
we conduct a hard selection on the optimal pruned transformers and skipping operation, which can be expressed as follows:
\begin{equation}
    \begin{aligned}
        &\boldsymbol{T}^{(l)} = \boldsymbol{T}_{k_1}^{(l)},\ \  where\ \  k_1 = \arg\max_i \alpha_{i},\\
        &L_{o} = L_{k_2},\ \ where \ \  k_2 = \arg\max_{j} \beta_j,\\
        &\forall l\in [1,L_o], \forall i\in[1,m], \forall j\in [1,L],
    \end{aligned}
\end{equation}
where $L_o$ is the optimal number of transformer blocks, $k_1$ and $k_2$ represent the decision for hard selection on candidate pruned transformers and skipping operations, respectively. We stack the candidate transformer $\boldsymbol{T}^{(l)}$ into $L_o$ layers to form the reconstructed network, which is also displayed in Figure \ref{fig:network}.
Before retraining model parameters, we transfer the well-trained model weights within the selected architecture from $\boldsymbol{\mathcal{W}}^{\ast}$ to parameter initialization of the corresponding architecture in the retraining stage. After that, the training dataset will be fed into the reconstructed model to generate predictions. Then all model weights will be updated by back-propagation.
The retrained model shortens the inference time and retains the accuracy of recommendation simultaneously.

%% file: Content/EXPERIMENT.tex
\begin{table}[t]
    \caption{\textbf{Statistical Information of Adopted Datasets}}
    \label{tab:data}
    \renewcommand{\arraystretch}{0.8}
    \resizebox{0.95\linewidth}{!}{
    \begin{tabular}{cccccc}
        \toprule
        Datasets& \# Users & \# Items & UA (AVE) & IA (AVE) &\# Interactions\\
        \midrule
        ML-1M & 6041 & 3707 & 165.60 & 269.89 & 1,000,209\\
        Gowalla & 64,116 & 164,533 & 31.48 & 12.27 & 2,018,421\\
        Douban & 110,091 & 29 & 10.55 & 41,496.71 & 2,125,056\\
        \bottomrule
    \end{tabular}}
    \vspace{-5mm}
\end{table}
\section{EXPERIMENTS}
    In this section, we perform four extensive experiments to show the effectiveness of our DNS-Rec Framework. First, we introduce our experimental settings, and then experiment results will be demonstrated and analyzed. Specifically, our experiments are set to answer these research questions:
    \vspace{-\topsep}
    \begin{itemize}[leftmargin=*]
        \item \textbf{RQ1:} How does the searched network structure perform?
        \item \textbf{RQ2:} To what extent does the searched network reduce the inference time and storage consumption? 
        \item \textbf{RQ3:} How do different components contribute to DNS-Rec?
        \item \textbf{RQ4:} How do the tradeoff parameter $\lambda$, number of gate layers $L_d$ and learning rate $\xi$ affect the model performance?
    \end{itemize}
    \vspace{-\topsep}

    \subsection{Datasets and Evaluation Metrics}
    As shown in Table \ref{tab:data}, we evaluate proposed DNS-Rec based on three benchmark datasets ML-1M~\footnote{https://grouplens.org/datasets/movielens/}, Gowalla~\footnote{https://snap.stanford.edu/data/loc-gowalla.html} and Douban~\footnote{https://www.kaggle.com/datasets/utmhikari/doubanmovieshortcomments}.

    Evaluation metrics, including Recall, Mean Reciprocal Rank (MRR), and Normalized Discounted Cumulative Gain (NDCG), are used for our experiments. Consistent with previous research~\cite{Kang01, News, Graph, Multi, Linrec}, the interactions are grouped by users chronologically, and the datasets are split by the leave-one-out strategy~\cite{Linrec}. To be more specific, the penultimate and the last item of the interaction sequence is used for validation and testing, respectively. Therefore, the ratio of the validation set to the test set is 1:1, and their sizes are determined by the number of users.
 
    \subsection{Baselines}
    To demonstrate the effectiveness and efficiency of our approach, we compare our approach with different state-of-the-art NAS methods and transformer-based models.

    \noindent\textbf{NAS Methods:} (1) \textbf{DARTS} ~\cite{DARTS}: relaxes the discrete search space into a continuous one, and searches the optimal architecture using gradient descent. (2) \textbf{ProxylessNAS} ~\cite{Proxylessnas}: models resource constraint as a continuous value loss and applies it to the differentiable NAS framework, which results in a compact searched network. (3) \textbf{RecNAS} ~\cite{RecNAS}: applies annealing and channel pruning to the NAS process, and can efficiently find high-performance architectures under given resource constraints.

     \noindent\textbf{Transformer-based Models:} (1) \textbf{SASRec} ~\cite{Kang01}: employs the multi-head attention scheme within transformer layers, facilitating the capture of both long-/short-term interactions. (2) \textbf{FDSA} ~\cite{FDSA}: integrates sequences at both the item and feature levels to further exploit the sequential behaviors. (3) \textbf{SASRecF} \footnote{https://www.recbole.io/docs/index.html}: 
    jointly consider feature fields and item fields as contextual information, thereby enhancing the efficacy of the SASRec model. We also apply \textbf{LinRec} ~\cite{Linrec} to these baselines, which improves the efficiency of the transformer-based models by utilizing a special mapping.

    \begin{table}[t]
		\caption{\textbf{Hyperparameters settings for DNS-Rec}}
		\label{tab:implementation}
        \renewcommand{\arraystretch}{0.9}
        \resizebox{\linewidth}{!}{
		\begin{tabular}{cccc}
			\toprule
		  Hyperparameters & ML-1M & Gowalla & Douban \\
            \midrule 
            hidden size $d$ of supernet  & 128 & 64 & 128 \\
            inner size $D$ of supernet & 256 & 256 & 256\\
            length of sequence $N$  & 200 & 100 & 100\\
            $L$ TRFM layers  & 4 & 4 & 4 \\
            $n$ attention heads  & 4 & 4 & 4 \\
            tradeoff parameter $\lambda$ & 0.1 & 0.1 & 0.1 \\
            learning rate $\xi$ & 0.001 & 0.001 & 0.001\\
            Dropout rate & 0.2 & 0.2 & 0.2 \\
            batch size (search) & 1024 & 1024 & 1024\\
            batch size (retraining) & 2048 & 2048 & 2048\\
            pruning intensities $\gamma$ & \{0\%, 25\%, 50\%\} & \{0\%, 50\%, 75\%\} & \{0\%, 50\%, 75\%\}\\
            pruning intensities $\gamma'$ & \{0\%, 25\%, 50\%\} & \{0\%, 50\%, 75\%\} & \{0\%, 50\%, 75\%\}\\
            top-$k$ for evaluation & 10 & 10 & 10\\
            temperature $\tau$ in $t$-th iteration& \multicolumn{3}{c}{$\tau = \max(0.01, 1-0.00005\cdot t)$} \\
			\bottomrule
		\end{tabular}}
    \vspace{-5mm}
	\end{table}

    \begin{table*}
		\caption{\textbf{Overall performance comparison between DNS-Rec and baselines}}
		\label{tab:performance}
        \renewcommand{\arraystretch}{0.85}
        \resizebox{0.95\linewidth}{!}{
		\begin{tabular}{cc >{\centering\arraybackslash}p{1.75cm}>{\centering\arraybackslash}p{1.75cm}>{\centering\arraybackslash}p{1.75cm}>{\centering\arraybackslash}p{1.75cm}>{\centering\arraybackslash}p{1.75cm}>{\centering\arraybackslash}p{1.75cm}}
			\toprule
			\multirow{2}{*}{Datasets} & \multirow{2}{*}{Methods} & \multicolumn{3}{c}{Evaluation Set} & \multicolumn{3}{c}{Test Set}\\
            \cline{3-8}
            \vspace{-2.5mm}\\
            && Recall@10 & MRR@10 & NDCG@10 &Recall@10 & MRR@10 & NDCG@10\\
			\midrule
			\multirow{9}{*}{ML-1M} & Ours & \textbf{0.7152$^{\ast}$} & \textbf{0.4207} &  \textbf{0.4910$^{\ast}$}& 0.6796 & \textbf{0.3991$^{\ast}$} & \textbf{0.4661$^{\ast}$} \\
            & DARTS + SASRec & 0.7142 & 0.4097 & 0.4825 & \textbf{0.6846$^{\ast}$} & 0.3915 & 0.4614\\
            & ProxylessNAS + SASRec  & 0.7013 & 0.4025 & 0.4740 & 0.6775 & 0.3819 & 0.4525\\
            & RecNAS + SASRec  & 0.7094 & 0.4090 & 0.4809 & 0.6791 & 0.3852 & 0.4554\\
            & SASRec (w/o LinRec) & 0.7118 & 0.4108 & 0.4829 & 0.6793 & 0.3822 & 0.4533\\
            & FDSA (w/o LinRec) & 0.6970 & 0.3946 & 0.4668 & 0.6677 & 0.3747 & 0.4447\\
            & SASRecF (w/o LinRec) & 0.7144 & 0.4166 & 0.4877 & 0.6783 & 0.3941 & 0.4620\\
            & SASRec (w LinRec) & 0.7134 & 0.4192 & 0.4894 & 0.6790 & 0.3949 & 0.4627\\
            & FDSA (w LinRec) & 0.7046 & 0.4099 & 0.4803 & 0.6753 & 0.3921 & 0.4597\\
            & SASRecF (w LinRec) & 0.7099 & 0.4176 & 0.4876 & 0.6815 & 0.3978 & 0.4657\\
            \midrule
			\multirow{9}{*}{Gowalla} & Ours & \textbf{0.9274$^{\ast}$} & \textbf{0.6833$^{\ast}$} &  \textbf{0.7425$^{\ast}$} & 0.9199 & \textbf{0.6715$^{\ast}$} & \textbf{0.7317$^{\ast}$} \\
            & DARTS + SASRec & 0.9250 & 0.6694& 0.7315 & 0.9179 & 0.6588 & 0.7217\\
            & ProxylessNAS + SASRec & 0.9271 & 0.6552 & 0.7212 & \textbf{0.9218$^{\ast}$} & 0.6438 & 0.7112 \\
            & RecNAS + SASRec & 0.9223 & 0.6601 & 0.7237 & 0.9144 & 0.6494 & 0.7136 \\
            & SASRec (w/o LinRec)& 0.9241 & 0.6598 & 0.7238 & 0.9159 & 0.6468 & 0.7119 \\
            & FDSA (w/o LinRec)& 0.8990 & 0.6231 & 0.6898 & 0.8894 & 0.6079 & 0.6759 \\
            & SASRecF (w/o LinRec)& 0.9069 & 0.6332 & 0.6994 & 0.8015 & 0.6207 & 0.6885 \\
            & SASRec (w LinRec)& 0.9247 & 0.6818 & 0.7408 & 0.9174 & 0.6679 & 0.7284 \\
            & FDSA (w LinRec)& 0.8957 & 0.6169 & 0.6863 & 0.8878 & 0.6056 & 0.6737 \\
            & SASRecF (w LinRec) & 0.9069 & 0.6294 & 0.6965 & 0.8987 & 0.6165 & 0.6847 \\
            \midrule
			\multirow{9}{*}{Douban} & Ours & \textbf{0.8919$^{\ast}$} & \textbf{0.4879$^{\ast}$} &  \textbf{0.5851$^{\ast}$} & 0.7854 & \textbf{0.3886$^{\ast}$} & \textbf{0.4833$^{\ast}$}\\
            & DARTS + SASRec & 0.8831 & 0.4731 & 0.5713 & 0.7813 & 0.3762 & 0.4724 \\
            & ProxylessNAS + SASRec & 0.8854 & 0.4710 & 0.5701 & 0.7830 & 0.3787 & 0.4746 \\
            & RecNAS + SASRec & 0.8783 & 0.4720 & 0.5693 & 0.7888 & 0.3797 & 0.4773 \\
            & SASRec (w/o LinRec)& 0.8827 & 0.4820 & 0.5784 & 0.7889 & 0.3858 & 0.4822 \\
            & FDSA (w/o LinRec)& 0.8870& 0.4740 & 0.5729 & \textbf{0.8050$^{\ast}$} & 0.3825 & 0.4827 \\
            & SASRecF (w/o LinRec)& 0.8856 & 0.4775 & 0.5756 & 0.7625 & 0.3758 & 0.4684 \\
            & SASRec (w LinRec) & 0.8767 & 0.4697 & 0.5673 & 0.7757 & 0.3710 & 0.4673 \\
            & FDSA (w LinRec) & 0.8823 & 0.4757 & 0.5733 & 0.7813 & 0.3794 & 0.4751 \\
            & SASRecF (w LinRec) & 0.8859 & 0.4791 & 0.5769 & 0.7879 & 0.3780 & 0.4757 \\
			\bottomrule
		\end{tabular}}
        \begin{tablenotes}
            \small\centering
            \item[*] Results of backbone models with different architecture search methods, with (w) and without (w/o) LinRec mechanism have been shown. $``\ast"$ indicates the improvements are \textbf{statistically significant} (i.e., two-sided t-test with $p$ < 0.05) over baselines), except Recall@10 on test set.
        \end{tablenotes}
        \vspace{-3mm}
	\end{table*}
    \subsection{Implementation}
    In this subsection, we introduce the implementation details of the proposed DNS-Rec Framework, which are shown in Table \ref{tab:implementation}. Hyperparameters except for batch size, number of layers and hidden/inner size remain unchanged from the search stage to the retraining stage.
    
    It is noteworthy the parameter $N$ is set to be larger than the average of users' actions in Table \ref{tab:data} to make sure most of the actions could be considered. Additionally, the parameter initialization follows the suggestions of ~\cite{Kaiming, Xavier}. We optimize DNS-Rec during both the search and retraining stage with Adam optimizer~\cite{Adam}. As the architecture weights converge in only a few training steps, we adopt a special 10-epoch early stopping strategy for the search stage. Hyperparameter settings for transformer-based baselines follow the suggestion of previous work~\cite{Linrec}.

    \subsection{Overall Performance (RQ1)}
    In this subsection, we compare DNS-Rec with both state-of-the-art efficient gradient-based search methods and attention-based SRS models that use manually selected architectures. Our goal is to demonstrate DNS-Rec's efficiency and to illustrate the often underestimated potential for model compression. The results, as shown in Table~\ref{tab:performance}, highlight the best outcomes in bold.
    
    From these results, it is evident that our method surpasses nearly all baseline models across the three benchmark datasets. This achie\-vement underscores DNS-Rec's ability to effectively identify and tailor the most suitable network architectures for specific recommendation tasks, thereby enhancing model performance.
    Traditional NAS methods, such as DARTS, ProxylessNAS, and RecNAS, tend to fall short in their search for optimal networks. DARTS often over-focuses on architectural accuracy, leading to overly complex network structures. ProxylessNAS and RecNAS are capable of finding smaller networks~(as shown in Table~\ref{tab:consumption}), but suffer from decreased accuracy. In contrast, DNS-Rec maintains model accuracy more effectively, especially noticeable in the Gowalla dataset.
    Moreover, when compared with state-of-the-art models with manually selected architectures, DNS-Rec shows significant improvements. This suggests that the network structures of these models are not optimally adapted to specific recommendation tasks. While baseline models with linear and non-linear attention mechanisms perform well on standard evaluation metrics, architectures selected based on prior experience might potentially be over-parameterized.

    In summary, our analysis reveals the superiority of DNS-Rec in balancing model complexity with task-specific performance, indicating its potential for more efficient and effective SRSs.
    
    \subsection{Efficiency Comparison (RQ2)}
 
    In this subsection, we conduct a comparative analysis of model efficiency, particularly examining how DNS-Rec outperforms contemporary NAS methods and models in terms of inference time and GPU memory occupation under resource constraints. Additionally, we analyze the trade-offs between the efficiency gains during inference and the costs during the search phase.
    \subsubsection{Inference Efficiency}
    The results, presented in Table~\ref{tab:consumption}, provide several critical insights: Firstly, DNS-Rec facilitates the identification of more compact network architectures, thereby significantly reducing GPU memory occupation and shortening inference time. By contrast, DARTS struggles to search for a compact architecture and usually incurs extremely high inference costs, given its lack of consideration for resource constraints. Meanwhile, although ProxylessNAS and RecNAS reduce the complexity of the network architectures, our method outperforms both of them on the three datasets due to our elaborately designed efficient supernet. Additionally, standard baseline model architectures, often selected based on prior experience, are prone to being oversized, which results in extended inference times and excessive GPU memory usage. Compared to linear and non-linear attention-based models, DNS-Rec achieves up to an $87.6\%$ reduction in inference time and a significant $91.1\%$ decrease in GPU memory usage. This efficiency stems from our method's automated search on the architectures' width and depth.
    \subsubsection{Search \& Inference Trade-offs}
    Although the search space and the supernet size are large, our DNS-Rec can also search efficiently. As for the ML-1M dataset, the search speed of DNS-Rec is 814s/epoch, compared with 905s/epoch using the state-of-the-art NAS method ProxylessNAS. In most cases, the architectural weights of DNS-Rec can stabilize rapidly, indicating that in practical applications, we can quickly search for very precise results. In Table 12, we demonstrate the comparison of the efficiency trade-offs between DNS-Rec and manually constructed baselines on the dataset ML-1M and use all the well-trained models to infer another similar movie dataset ML-10M with 10M interactions. 

    \subsection{Ablation Study (RQ3)}
    In this subsection, we analyze the efficacy of the data-aware gate components in DNS-Rec, which is pivotal for maintaining accuracy. Each data-aware gate comprises two linear layers. To investigate its essential role in searching the architectures and enhancing the model performance in the inference stage, we sequentially remove these layers.  Focusing on the ML-1M dataset, the results outlined in Table~\ref{tab:ablation}, offer insightful observations.
    \begin{table}
		\caption{\textbf{Efficiency comparison}}
		\label{tab:consumption}
        \renewcommand{\arraystretch}{0.9}
        \resizebox{\linewidth}{!}{
		\begin{tabular}{cccc}
			\toprule
			Datasets & Model & Infer. & GPU Memory\\
			\midrule
			\multirow{10}{*}{ML-1M} &  Ours& \textbf{14.8s} & \textbf{7.78GB}\\
            & DARTS + SASRec& 23.0s & 16.31GB \\
            & ProxylessNAS + SASRec & 18.3s & 11.60GB\\
            & RecNAS + SASRec & 18.5s & 11.60GB\\
            & SASRec (w/o LinRec)& 54.2s & 28.26GB\\
            & FDSA (w/o LinRec)& 92.8s & 45.21GB\\
            & SASRecF (w/o LinRec)& 49.8s & 24.88GB\\
            & SASRec (w LinRec) & 30.3s & 19.47GB\\
            & FDSA (w LinRec) & 49.8s & 35.64GB\\
            & SASRecF (w LinRec) & 24.8s & 17.57GB\\
            \cline{2-4}
            \vspace{-2.8mm}\\
            & Imprv. & 1.9\%$\sim$84.1\% & 32.9\% $\sim$ 82.8\%\\
            \midrule
			\multirow{10}{*}{Gowalla} & Ours &  \textbf{232s} & \textbf{5.54GB}\\
            & DARTS + SASRec & 268s & 8.60GB \\
            & ProxylessNAS + SASRec& 237s & 6.39GB \\
            & RecNAS + SASRec& 234s & 7.40GB \\
            & SASRec (w/o LinRec)& 383s & 13.75GB\\
            & FDSA (w/o LinRec) & 565s & 21.68GB\\
            & SASRecF (w/o LinRec)& 385s & 15.38GB\\
            & SASRec (w LinRec) & 393s & 10.44GB\\
            & FDSA (w LinRec) & 420s & 16.66GB\\
            & SASRecF (w LinRec) & 321s & 10.79GB\\
            \cline{2-4}
            \vspace{-2.8mm}\\
            & Imprv. & 0.9\%$\sim$58.9\% & 13.3\% $\sim$ 74.4\%\\
            \midrule
			\multirow{10}{*}{Douban} & Ours & \textbf{70s} & \textbf{1.20GB}\\
            & DARTS + SASRec& 242s & 7.11GB\\
            & ProxylessNAS + SASRec & 109s & 1.82GB\\
            & RecNAS + SASRec & 82s & 1.76GB\\
            & SASRec (w/o LinRec)& 357s & 9.42GB\\
            & FDSA (w/o LinRec)& 563s & 13.45GB\\
            & SASRecF (w/o LinRec)& 365s & 9.62GB\\
            & SASRec (w LinRec) & 223s & 6.86GB\\
            & FDSA (w LinRec) &  369s & 9.90GB\\
            & SASRecF (w LinRec) & 201s & 8.30GB\\
            \cline{2-4}
            \vspace{-2.8mm}\\
            & Imprv. & 14.6\%$\sim$87.6\% & 31.8\% $\sim$ 91.1\%\\
			\bottomrule
		\end{tabular}}
        \vspace{-5mm}
	\end{table}
    \begin{table}
		\caption{\textbf{The trade-offs between the efficiency gains during the inference stage and the costs during the search stage.}}
		\label{tab:tradeoff}
        \renewcommand{\arraystretch}{1}
        \resizebox{\linewidth}{!}{
		\begin{tabular}{ccccc}
			\toprule
			Methods & Training$^{\ast}$ & Infer. (Eval) & Infer. (10M Inters) & Total\\
			\midrule
			DNS-Rec & 234min & 14.8s & 408min & \textbf{642min} \\
            SASRec(w LinRec) & 126min & 30.3s & 836min & 962min \\
            SASRec(w/o LinRec) & 174min & 54.2s & 1495min & 1669min \\
			\bottomrule
		\end{tabular}}
        \begin{tablenotes}
            \small\centering
            \item[*] $``\ast:"$ The training time includes the time cost of the search stage and the retraining stage. The search time of manually constructed baseline models SASRec (w or w/o LinRec) is 0. 
        \end{tablenotes}
        \vspace{-5mm}
	\end{table}
    We design two types of ablation studies to verify the efficacy of the data-aware gates. For the first type named "Retrain Only", we employ 2-layer gates to search an appropriate network fitted to the 2-layer gates, and then remove the linear layers sequentially in the retraining stage. For the second type named "Search \& Retrain", we remove the linear layers in both the search and inference stages. 
    
    The data indicates that the incorporation of two layers in each data-aware gate substantially enhances accuracy compared with using one linear layer or removing the gate entirely. The performance enhancement shows that the data-aware gates are able to perceive the input data environment and retain the accuracy for smaller model architectures. Specifically, our method, with the fully implemented gate in the retraining stage, shows accuracy improvements of 2.43$\%$, 8.57$\%$, and 6.39$\%$ on Recall@10, MRR@10, and NDCG@10, respectively, compared with architectures without gates. 

   To further demonstrate the effectiveness of data-aware gates in the NAS process, we conduct another ablation study shown in Table \ref{fig:gate} where the gates are ablated in both the search and retraining stages. If we remove one linear layer from the gates in both the search and retraining stages, DNS-Rec with single-layer gates will generate searched networks with single-layer gates. As the data-aware gates are ablated in both stages, the search results may be different. These results verify the efficacy of the data-aware gates and the gates enable the architectures to be searched based on the data environment, which effectively improves the NAS process.
   
    In summary, the performance enhancement contributes to our data-aware gates that dynamically generate the networks by adapting to the perception process of input data batch. These findings not only verify the data-aware gates' effectiveness but also underscore the significance of leveraging the data perception process.

    \begin{table}
		\caption{\textbf{Ablation study of data-aware gates}}
		\label{tab:ablation}
        \renewcommand{\arraystretch}{0.9}
        \resizebox{\linewidth}{!}{
		\begin{tabular}{ccccc}
			\toprule
			Type & Component & Recall@10 & MRR@10 & NDCG@10\\
			\midrule
			\multirow{4}{*}{Retrain Only} & 2 layers & \textbf{0.7151$^{\ast}$} & \textbf{0.4207$^{\ast}$} & \textbf{0.4910$^{\ast}$}\\
            & 1 layer & 0.7043 & 0.4059 & 0.4771\\
            & No Gates & 0.6977 & 0.3875 & 0.4615\\
            \cline{2-5}
            \vspace{-2.8mm}\\
            & Imprv. & 2.43\% & 8.57\% & 6.39\%\\
            \midrule
            \multirow{4}{*}{Search \& Retrain} & 2 layers & \textbf{0.7151$^{\ast}$} & \textbf{0.4207$^{\ast}$} & \textbf{0.4910$^{\ast}$}\\
            & 1 layer & 0.7131 & 0.4169 & 0.4876\\
            & No Gates & 0.7089 & 0.4141 & 0.4846\\
            \cline{2-5}
            \vspace{-2.8mm}\\
            & Imprv. & 0.87\% & 1.59\% & 1.32\%\\
			\bottomrule
		\end{tabular}}
        \begin{tablenotes}
            \small\centering
            \item[*] $``\ast"$ indicates the improvements are \textbf{statistically significant} (i.e., two-sided t-test with $p$ < 0.05) over baselines)
        \end{tablenotes}
        \vspace{-5mm}
	\end{table}
 
    \subsection{Parameter Analysis (RQ4)}
    In this subsection, we explore the significant impact of the tradeoff parameter $\lambda$ on architecture search within the supernet, particularly under varied resource constraints. This dynamic parameter is pivotal in modulating resource constraint intensity, crucially influencing the balance between system performance and resource efficiency in our recommender system. In addition, we discuss the sensitivity of the model's performance to the number of layers $L_d$ in the data-aware gates and the learning rate $\xi$. We present an in-depth analysis of how varying values of $\lambda$, $L_d$, and $\xi$ affect different evaluation metrics and FLOPs providing essential insights into the model's adaptability to diverse application scenarios. 
    
    \textbf{Tradeoff parameter $\lambda$}. As depicted in Figure ~\ref{fig:parameter}, our findings reveal a direct correlation between $\lambda$ and the network's structural compactness: as $\lambda$ increases, the penalty intensity increases correspondingly, which leads to more compact networks with reduced computational costs. Selections on different values of parameter $\lambda$ result in comparable performance, illustrating the robustness and rationality of our search method under dynamic resource constraints.
    Furthermore, the observation that different $\lambda$ values may result in identical architectures offers a flexible range of choices in model configuration. For instance, a smaller $\lambda$ is optimal and provides satisfactory performance when resources are plentiful, while a larger $\lambda$ is more suited to real applications under limited resources, effectively minimizing the computational cost. Overall, the tradeoff parameter $\lambda$ enables the network architectures to adapt flexibly to various practical applications, demonstrating robust performance in recommender systems despite limited resources.

    \textbf{Number of Layers $L_d$ in the Data-aware Gates}. We change the number of layers $L_d$ in the data-aware gates throughout the search stage and the retraining stage. The model performance and the corresponding FLOPs are shown in Table \ref{tab:l_d}. We can find that data-aware gates with 2 to 3 layers are the most effective and efficient. Adding too many layers to the data-aware gates might lead to inappropriate search results and performance drops during the model inference. As we add the gate layers, the FLOPs in the retraining stage increase correspondingly, because the large gates increase the depth of the model substantially.

    \textbf{Learning Rate $\xi$}. We also change $\xi$ in [0.0001, 0.0005, 0.001, 0.003, 0.005, 0.01] to verify the robustness of the search process. We found that changing the learning rate of the search stage within a reasonable range does not result in different search results, which further indicates the effectiveness and stability of DNS-Rec.

%% file: Content/Related_works.tex
 \begin{figure}[t]
		\centering
		\includegraphics[width=0.95\linewidth]{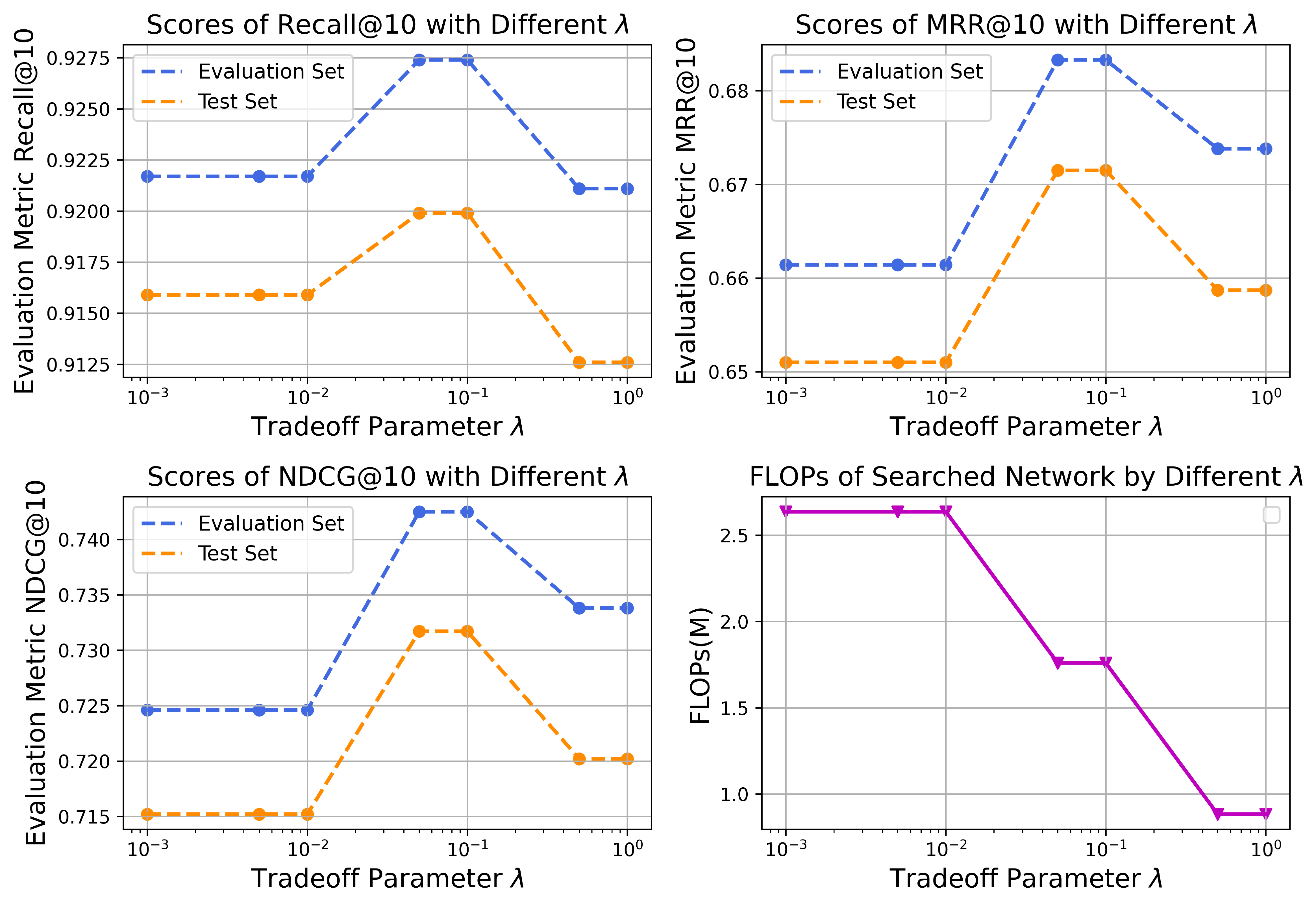}
		\caption{\textbf{Impacts of different tradeoff parameters $\lambda$ on evaluation metrics and FLOPs}}
		\label{fig:parameter}
    \vspace{-5mm}
	\end{figure}
\section{Related Works}
\noindent\textbf{Transformer-based SRSs}
Diverse studies have been conducted to leverage transformer blocks to learn sequence representations from users, as the attention mechanism in the transformer can effectively capture the historical interactions. Models like SASRec ~\cite{Kang01} employ transformer layers to make predictions based on long-/short-term interactions.
FDSA~\cite{FDSA} combines item-level and feature-level sequences together to enhance the extraction of sequential patterns. 
TSCR ~\cite{TSCR} models item sequence and contextual information from conversations together to achieve better recommendation performance. AdaMCT ~\cite{ADAMCT} incorporates convolution neural networks into the transformer-based recommender system to model historical interactions. Uniquely, our method, Data-aware Neural Architecture Search for Recommender Systems (DNS-Rec) pioneers the realm of automatically exploring smaller yet more efficient architectures for transformer-based SRSs.

\noindent\textbf{Resource Constrained NAS}
The evolution of resource-aware NAS techniques marks a significant shift in balancing performance with computational efficiency in recommender systems by integrating computational constraints.
Models like MONAS ~\cite{hsu2018monas} utilize policy-based reinforcement learning, incorporating operation quantities directly into the reward function. Differentiable NAS~\cite{DARTS} frameworks often employ metrics such as FLOPs, parameter size, and latency as penalty terms in the loss function, exemplified by ProxylessNAS ~\cite{Proxylessnas} and FBNet~\cite{wu2019fbnet}. ProxylessNAS models the architecture's latency as a continuous value loss, while FBNet uses a lookup table for latency estimation of candidate operations. SNAS~\cite{xie2018snas} emphasizes the differentiability of resource constraints, modeling network costs through linear functions.
In contrast, our method, the Data-aware Neural Architecture Search for Recommender Systems (DNS-Rec), advances this field by focusing on recommendation accuracy while adhering to resource constraints, setting a new standard in the efficiency and effectiveness of recommender systems.

%% file: Content/conclusion.tex
\begin{table}
		\caption{\textbf{Impacts of the Layer Number $L_d$ in the Data-aware Gates on the evaluation metrics and FLOPs}}
		\label{tab:l_d}
        \renewcommand{\arraystretch}{0.8}
        \resizebox{0.75\linewidth}{!}{
		\begin{tabular}{ccccc}
			\toprule
			$L_d$ & Recall@10 & MRR@10 & NDCG@10 & FLOPs\\
            \midrule
            0 & 0.7089 & 0.4141 & 0.4846 & 20M\\
            1 & 0.7131 & 0.4169 & 0.4876 & 27M\\
            2 & 0.7151 & 0.4207 & 0.4910 & 26M\\
            3 & 0.7123 & 0.4238 & 0.4929 & 35M\\
            4 & 0.7121 & 0.4224 & 0.4916 & 64M\\
			\bottomrule
		\end{tabular}}
        \vspace{-5mm}
	\end{table}

\section{Conclusion}
In this paper, we introduced an innovative approach to reduce computational overhead and resource inefficiency in Sequential Recommender Systems (SRSs) using attention-based mechanisms. Our Data-aware Neural Architecture Search for Recommender Systems (DNS-Rec) enhances both efficiency and accuracy. DNS-Rec integrates a data-aware gate mechanism and a dynamic resource constraint strategy, effectively reducing computational demands while improving recommendation performance. This method adapts historical user-item interactions into deep transformer layers, showing substantial promise in our extensive testing across various datasets. The results confirm DNS-Rec’s superiority over existing models and underscore its potential to enhance recommendation system efficiency without sacrificing accuracy. The principles developed here have broader implications, potentially impacting various machine learning and artificial intelligence applications, and providing a foundation for future research in efficient recommender systems.